\documentclass[aps,twocolumn,noeprint]{revtex4-1}
\usepackage{graphicx}
\usepackage{dcolumn} 
\usepackage{bm}
\usepackage{amsmath}
\usepackage{multirow}
\usepackage{xcolor}
\begin{document}
\title{Nuclear shell model study of neutrinoless double beta decay under Left-Right symmetric model}
\author{Dong-Liang Fang$^{a,b}$, B. Alex Brown$^{c}$ and Fedor \v{S}imkovic$^{d,e}$} 
\address{ $^a$Institute of Modern Physics, Chinese Academy of sciences, Lanzhou, 730000, China}
\address{ $^b$University of Chinese Academy of Sciences, Beijing, 100049,China}
\address{ $^c$Department of Physics and Astronomy, FRIB Laboratory, Michigan State University, East Lansing, MI 42284-1321, USA}
\address{ $^d$Faculty of Mathematics, Physics and Informatics, Comenius University in Bratislava, 842 48 Bratislava, Slovakia}
\address{$^e$Institute of Experimental and Applied Physics, Czech Technical University in Prague, 110 00 Prague, Czech Republic}
\begin{abstract}
We use the large scale nuclear shell model to calculate the nuclear matrix elements for the neutrino mediated neutrinoless double beta decay within the Left-Right symmetric model for four nuclei: $^{76}$Ge, $^{82}$Se, $^{130}$Te and $^{136}$Xe. We perform a systematic analysis on the general magnitude of different terms for related mechanisms. For the $\eta$ mechanism, we find that the weak magnetism $R$ term dominates the decay rate while the $p$-wave effect is suppressed. While for the $\lambda$ mechanism, the $\omega$ and the $q$ terms are with equal importance. For the latter $q$ term, important contributions from weak-magnetism MM part are observed. Finally, we give the constraints on the new physics parameters $m_{\beta\beta}$, $\lambda$ and $\eta$ from current experiments.
\end{abstract}
\maketitle
\section{Introduction}
The neutrinoless double beta ($0\nu\beta\beta$) decay is one of the rarest and most mysterious processes in nature. With the construction of the deep-underground laboratories and advancements of many up-to-ton-scale experiments \cite{CDEX:2023owy,NvDEx:2023zht,PandaX-II:2019euf,nEXO:2021ujk}, we are approaching the so-called inverted mass hierarchy region \cite{Agostini:2022zub}. In the next few decades, we may see the signal of this lepton-number violating process and answer the questions of whether the neutrino is a Majorana particle and subsequently why the universe is dominated by matter instead of antimatter. 

The discovery of neutrinoless double beta decay will pave our way to physics beyond the standard model (SM). While this discovery will prove lepton number violation, we still need to understand the details of this process and what is actually happening, whether it is mediated by the simple light neutrino with SM weak currents or with more complicated weak currents. These are closely related to new physics models. One of the promising models of this kind is the so-called Left-Right symmetric model \cite{Mohapatra:1980yp}. For this model, the see-saw mechanism can be naturally incorporated \cite{Mohapatra:2005wg}, addressing the issue of the smallness of neutrino mass. Such studies have already attracted much attention from the community, and the implications for future observations of $0\nu\beta\beta$-decay have been intensively discussed \cite{Tello:2010am,Chakrabortty:2012mh,Awasthi:2013ff,Deppisch:2014zta,Pritimita:2016fgr,Fukuyama:2022naj}.  

Another issue is that if this model turns out to be the naturally chosen one, how could we confirm this? Various approaches have been proposed. For example, one could compare the decay half-lives of different isotopes by taking advantage of the different NME dependencies on various mechanisms for different nuclei \cite{Simkovic:2010ka,Faessler:2011qw,Faessler:2011rv,Meroni:2012qf,Graf:2022lhj,Agostini:2022bjh}. One could measure the angular correlation of the two emitted electrons \cite{Graf:2022lhj}. One could also resort to comparing the different decay modes of the same isotope \cite{Simkovic:2001ft,Fang:2022cos}.

While most discussions focus on the new physics side, it is important to note that calculations from the nuclear side, which are crucial for our understanding of this process, are still lacking. Compared to the extensive nuclear many-body calculations for $0\nu\beta\beta$-decay with a standard neutrino mass mechanism \cite{Yao:2021wst,Agostini:2022zub}, the nuclear structural calculation for the LR symmetric model related to the $\lambda$ and $\eta$ mechanisms are limited and incomplete. Earlier publications have utilized methods such as large scale shell models (LSSM) \cite{Caurier:1996bu}, quasi-particle random phase approximation (QRPA) \cite{Muto:1989cd,Pantis:1996py,Suhonen:1998ck} and projected Hatree-Fock Boglyubov (PHFB) \cite{Tomoda:1990rs}. However, these studies have omitted the important leading order (LO) component, such as the Pseudo-Scalar component of the nuclear current. Subsequent calculations \cite{Simkovic:2017uvu,Sarkar:2020cjn,Iwata:2021bgi} still lack a thorough investigation of all the relevant terms.

Especifically for the LSSM calculations, we have abundant calculations of various nuclei for a standard neutrino mass mechanism, some with traditional Hamiltonians obtained by reproducing the nuclear properties (for a complete list, see {\it e.g.} \cite{Horoi:2022bru,Coraggio:2020iht}), while others are the so-called {\it ab initio} calculations \cite{Belley:2020ejd,Belley:2023btr} using the renormalized interactions derived from bare nucleon-nucleon interactions adjusted to produce nucleon scattering phase shifts. However, there are discrepancies among these calculations especially between the conventional and {\it ab initio} ones. This is not the topic of current study. For the mechanisms related to LR symmetric models, such calculations are limited \cite{Caurier:1996bu,Sarkar:2020cjn,Iwata:2021bgi,Horoi:2018fls}. Systematic studies of $0\nu\beta\beta$ with the LR symmetric model including comprehensive comparisons across various models and detailed error analysis, are still lacking.

In this work, we carry out LSSM calculations with traditional Hamiltonians to explore the long-range mechanism mediated by neutrinos in the LR symmetric model, incorporating key components of nuclear currents. We give a detailed analysis of the relative magnitude of each term associated with each mechanism and discuss how this difference originates from the nuclear side.

This article is arranged as follows, first the mechanism and related expression for the decay width, and then we give the NME results and some analysis of relative importance of each term, finally we give a brief discussion on how current results could be used to constrain the new physics parameters.

\section{$0\nu\beta\beta$ under LR symmetric model}
Starting from the basic LR symmetric model $SU(2)_L \otimes SU(2)_R \otimes U(1)_{B-L}$, one has extra gauge bosons $W_R$, $Z_R$, the triplet and the bi-doublet Higgs bosons $\phi$ and $H_R$, as well as right handed neutrinos $\nu_R$. In such a model, besides the electro-weak breaking energy scale $v_L$, there is another symmetry breaking energy scale $v_R$ which determines the general mass of right-handed gauge bosons. The Yukawa coupling between the neutrinos and Higgs bosons gives rise to the neutrino mass, and the see-saw mechanism can be naturally realized \cite{Mohapatra:1980yp}.

In such a model, the weak and the mass eigenstates of the gauge bosons are different, this then leads to the mixing between the left- and right-handed gauge bosons:
\begin{eqnarray}
\left(
\begin{array}{c}
     W_L  \\
     W_R
\end{array}
\right) = 
\left(
\begin{array}{cc}
    \cos\zeta & \sin\zeta  \\
     -\sin\zeta & \cos\zeta
\end{array}
\right)
\left(
\begin{array}{c}
     W_1  \\
     W_2
\end{array}
\right)
\end{eqnarray}
$W_L$ and $W_R$ are the gauge eigenstates while $W_1$ and $W_2$ are the mass eigenstates respectively. The masses for $W_1$ and $W_2$ Bosons are $M_1$ and $M_2$ respectively. And $\zeta$ is the mixing angle between the mass and weak eigenstates of weak bosons.

The interaction of the lepton and nucleon currents with $W$ Bosons can be written as \cite{Doi:1985dx,Tomoda:1990rs}:
\begin{eqnarray}
\mathcal{L}_{CC}=\frac{g}{2\sqrt{2}}[(j_L^{\mu}+J_L^{\mu})W^-_{L\mu}+(j_R^{\mu}+J_R^{\mu})W^-_{R\mu}]+h.c.
\end{eqnarray}
The lepton currents have the form~\cite{Tomoda:1990rs}:
\begin{eqnarray}
j_{L(R)}^\mu = \bar{e}_{L(R)}\gamma^\mu \nu_{eL(R)}
\end{eqnarray}

The Yukawa coupling of the neutrinos with the triplet Higgs Boson leads to the mixing between the left- and the right-handed neutrinos, but the Pontecorvo-Maki-Nakagawa-Sakata (PMNS) matrix \cite{Maki:1962mu,Pontecorvo:1967fh} is still with unknown origin:
\begin{eqnarray}
\nu_{eL}&=&\sum_{j=1}^{3} ( U_{ej} \nu_{j}+S_{ei} N_j^C ) \nonumber \\
\nu_{eR}&=&\sum_{j=1}^{3} (T^*_{ej} \nu_{j}^C + V_{ej}^* N_j)
\end{eqnarray}
Here $\nu_j$ and $N_j$ are the mass eigenstates of the light and the heavy neutrinos respectively. And U, S, T and V constitute a generalized 6$\times$6 PMNS matrix \cite{Xing:2011ur}.

The nucleon currents can be obtained by the Lorentz invariance with the impulse approximation \cite{Doi:1985dx,Simkovic:1999re}:
\begin{eqnarray}
J_{L(R)}^{\mu}(\vec{x})&=&\bar{p}(\vec{x})(g_V\gamma^\mu-ig_W\sigma^{\mu\nu}q_\nu \nonumber \\
&&\mp g_A\gamma^\mu \gamma_5 \pm g_P\gamma_5 q^\mu) n(\vec{x}) \nonumber 
\end{eqnarray}
The form factors $g_V$, $g_A$, $g_P$ and $g_W$ have the form given in \cite{Tomoda:1990rs}, and when reduced to the non-relativistic form, we have weak magnetism coupling constant $g_M(q^2)=g_V(q^2)+g_W(q^2)$. 

By integrating out the heavy gauge boson, we obtain the current-current interactions for these weak interactions as:
\begin{eqnarray}
H_{\rm int}&=&\frac{G\cos \theta_C}{\sqrt{2}}(j_{L\mu}J_L^{\mu\dagger}+\kappa j_{L\mu}J_R^{\mu\dagger} + \eta j_{R\mu}J_L^{\mu\dagger} \nonumber \\
&&+\lambda j_{R\mu}J_R^{\mu\dagger} ) + h.c.
\end{eqnarray}
Here the parameters $\lambda$, $\eta$ and $\kappa$ are combinations of parameters of the LR-symmetric model:
\begin{eqnarray}
\frac{G}{\sqrt{2}}&=&\frac{g^2}{8 M_1^2} (\cos^2 \zeta + (M_1/M_2)^2\sin^2\zeta) \approx \frac{g^2}{8M_1^2} \nonumber \\
\lambda &=& \frac{(M_{1}/M_{2})^2+\tan^2 \zeta}{1+(M_{1}/M_{2})^2\tan^2\zeta}\approx (M_1/M_2)^2+\tan^2 \zeta \nonumber \\
\eta&=&\kappa=-\frac{(1-(M_1/M_2)^2)\tan\zeta}{1+(M_1/M_2)^2\tan^2\zeta} \approx -\tan \zeta
\end{eqnarray}
If $\lambda$ has the same order as $\eta$, then the $\tan^2 \zeta$ term in $\lambda$ can be safely neglected, on the other hand if the first and second term in the r.h.s. of $\lambda$ have a  similar size ($M_{W1}/M_{W2}\sim \tan \zeta$), then the contribution from $\lambda$ to $0\nu\beta\beta$-decay can be safely neglected.
 
To better understand how the nuclear part and the lepton part contribute to the reaction matrix $R$ (the definition see Appendix C of \cite{Doi:1985dx}), we rewrite the decay width in \cite{Tomoda:1990rs, Stefanik:2015du} in the form:
\begin{eqnarray}
[T_{1/2}^{0\nu}]^{-1}&=&\mu_{\beta\beta}^2\mathcal{C}_{mm}+\mu_{\beta\beta}\langle \lambda \rangle \cos\psi_1 \mathcal{C}_{m\lambda}\nonumber \\
&+&\mu_{\beta\beta}\langle \eta \rangle \cos\psi_2 \mathcal{C}_{m\eta} 
+\langle \lambda \rangle^2 \mathcal{C}_{\lambda\lambda} \nonumber \\
&+& \langle \eta \rangle^2 \mathcal{C}_{\eta\eta}+\langle\lambda\rangle \langle\eta\rangle \cos(\psi_1-\psi_2) \mathcal{C}_{\lambda\eta}
\label{LRDW}
\end{eqnarray}
The new physics parameters directly related to the decay are $\mu_{\beta\beta}=m_{\beta\beta}/m_e$ ($m_{\beta\beta}=|\sum_{j} U^2_{ej} m_j|$ is the effective neutrino mass defined in the literature~\cite{Simkovic:1999re}), $\langle \lambda \rangle = |\lambda \sum_{j} U_{ej} V_{ej}|$ and $\langle \eta \rangle = |\eta \sum_{j} U_{ej} V_{ej}|$, the two phase angles have the form $\psi_1=arg[(\sum_j m_j U_{ej}^2)(\sum_{j}U_{ej}T_{ej}^*(g_V/g_V'))^*]$ and $\psi_2=arg[(\sum_j m_j U_{ej}^2)(\sum_{j}U_{ej}T_{ej}^*)^*]$. By assuming $g_V=g_V'$, we have $\psi_1\approx\psi_2$. 

The coefficients $\mathcal{C}$'s have the forms \cite{Stefanik:2015du}:
\begin{eqnarray}
\mathcal{C}_{mm}&=&\mathcal{G}_{01} |M^{0\nu}_m|^2 \nonumber \\
\mathcal{C}_{m\lambda}&=& -\mathcal{G}_{03} M^{0\nu}_m M^{0\nu}_{\omega-} +\mathcal{G}_{04} M^{0\nu}_m M^{0\nu}_{q+}  \nonumber \\
\mathcal{C}_{m\eta}&=& \mathcal{G}_{03} M^{0\nu}_m M^{0\nu}_{\omega+} - \mathcal{G}_{04} M^{0\nu}_m M^{0\nu}_{q-} - \mathcal{G}_{05} M^{0\nu}_m M^{0\nu}_{P} \nonumber \\
&+& \mathcal{G}_{06} M^{0\nu}_m M^{0\nu}_{R}\nonumber \\
\mathcal{C}_{\lambda\lambda}&=& \mathcal{G}_{02}|M^{0\nu}_{\omega-}|^2+\mathcal{G}_{011}|M^{0\nu}_{q+}|^2 \nonumber \\
\mathcal{C}_{\eta\eta}&=& \mathcal{G}_{02}|M^{0\nu}_{\omega+}|^2+\mathcal{G}_{011}|M^{0\nu}_{q-}|^2+ \mathcal{G}_{08}|M^{0\nu}_{P}|^2 \nonumber \\
&+&\mathcal{G}_{09}|M^{0\nu}_{R}|^2-\mathcal{G}_{07} M^{0\nu}_P M^{0\nu}_R\nonumber \\
\mathcal{C}_{\lambda \eta} &=& -2 \mathcal{G}_{02} M^{0\nu}_{\omega-}M^{0\nu}_{\omega+} -\mathcal{G}_{010}(M^{0\nu}_{q+}M^{0\nu}_{\omega+}+M^{0\nu}_{q-}M^{0\nu}_{\omega-}) \nonumber\\
&-& 2\mathcal{G}_{011} M^{0\nu}_{q+}M^{0\nu}_{q-} 
\end{eqnarray}

Here we use a different form of phase space factors from that of \cite{Stefanik:2015du} to better separate the effects of the $s$ and $p$-wave electrons. Their correspondence to the phase space factors of \cite{Stefanik:2015du} is as following: 
\begin{eqnarray}
\mathcal{G}_{01,02,03} &=& G_{01,02,03} \nonumber \\
\mathcal{G}_{04} &=& (3 m_e R) G_{04} \nonumber \\
\mathcal{G}_{05,06} &=& (m_e R) G_{05,06} \nonumber \\
\mathcal{G}_{07,08,09} &=& (m_e R)^2 G_{07,08,09} \nonumber \\
\mathcal{G}_{10} &=& 3 (m_e R)^2 \tilde{G}_{010} \nonumber \\
\mathcal{G}_{11} &=& (3m_e R)^2 \tilde{G}_{011} \nonumber
\end{eqnarray}

Meanwhile, the NMEs can be written as sums from different individual NMEs:
\begin{eqnarray}
M^{0\nu}_{m} &=& -M_F + M_{GT} + M_{T} \nonumber \\
M^{0\nu}_{\omega\pm} &=& M_{\omega GT\pm} + M_{\omega T\pm} \pm M_{\omega F} \nonumber \\
M^{0\nu}_{q \pm} &=& \frac{1}{3 m_e R} (M_{q GT\pm} -6 M_{q T\pm} \pm 3 M_{q F}) \nonumber \\
M^{0\nu}_{R} &=& \frac{1}{m_e R}(M_{R GT} + M_{R T}) \nonumber \\
M^{0\nu}_{P} &=& \frac{1}{m_e R} M_P
\label{indNME}
\end{eqnarray}

We notice that the reaction matrix has the form $R\sim M f_e(\epsilon_1,\epsilon_2)$ \cite{Doi:1985dx}, where M contains the information of the neutrino potential as well as nuclear transition amplitude, while $f_e$ is a functional of electron wave functions. Then, for the diagonal terms such as $\mathcal{C}_{mm}$, it is reasonable that $R\sim \mathcal{R}\equiv M \sqrt{\mathcal{G}}$. This newly defined term $\mathcal{R}$ can therefore be used for the comparison of general magnitude of components from various mechanisms, as we shall show in subsequent sections.

The individual NMEs $M_I$ can be written in a general form \cite{Stefanik:2015du} :
\begin{eqnarray}
M_{I}&=&\langle f || h_{I}(r,r_+)\mathcal{O}_I ||i\rangle \nonumber\\
&=&\sum_{p_1p_2 n_1 n_2 J}\langle f|| [c_{p1}^\dagger c_{p2}^\dagger]_{J} [\tilde{c}_{n2} \tilde{c}_{n1}]_J ||i\rangle \nonumber\\
&\times&\langle p_1 p_2 J||h_{I}(r,r_+)\mathcal{O}_I||n_1 n_2 J\rangle
\label{NME}
\end{eqnarray}

In such a treatment, the input from nuclear many-body approaches are the so-called two-body transition densities(TBTD) $\langle f|| [c_{p1}^\dagger c_{p2}^\dagger]_{J} [\tilde{c}_{n2} \tilde{c}_{n1}]_J ||i\rangle$. For our calculations, the TBTDs are obtained from the {\bf NuShellX} code \cite{Brown:2014bhl}.

The neutrino expotential $h_{I}(r,r+)$ can be written for the $m$, $q$, $R$ and $P$ terms in a general form as:
\begin{eqnarray}
h_{Ik}(r,r_+)=f_{src}^2(r)\sum_{k}\frac{2R}{\pi}\int f_{Ik}(q,r,r_{+})\frac{qdq}{q+\tilde{A}_m}
\end{eqnarray}
Here $k$ refers to contributions from different components of the weak current. {A closure energy $\tilde{A}_m$ is introduced to average the excitation energies from the intermediate states.} And $\vec{r}=\vec{r}_m-\vec{r}_n$ and $\vec{r}_+=(\vec{r}_m+\vec{r}_n)/2$ are the relative distance and the center of mass coordinate of the two decaying nucleons respectively. The short range correlations (src's) between the two nucleons is taken into account by the radial function $f_{src}(r)$. In our calculation, we adopt the CD-Bonn src \cite{Simkovic:2007vu}.

The radial functions $f_{I}$ in above $h_I$ have the form [with the usual convention  \cite{Stefanik:2015du}, where a factor $g_A(0)$ is taken out]  for the mass term:
\begin{eqnarray}
f_{F}&=&j_0(qr) g_V^2(q^2) \nonumber \\
f_{GT}&=& f_{GT,AA}+f_{GT,AP}+f_{GT,PP}+f_{GT,MM} \nonumber \\
&=&j_0(qr)[g_A^2(q^2)-\frac{2}{3}g_A(q^2)g_P(q^2)\frac{q^2}{2m_N}\nonumber \\
&+& \frac{1}{3} g_P^2(q^2) \frac{q^4}{4m_N^2} + \frac{2}{3}g_M^2(q^2) \frac{q^2}{4m_N^2} ] \nonumber \\
f_{T}&=& f_{T,AP}+f_{T,PP}+f_{T,MM} \nonumber \\
&=&j_2(qr)[\frac{2}{3}g_A(q^2)g_P(q^2)\frac{q^2}{2m_N} - \frac{1}{3} g_P^2(q^2) \frac{q^4}{4m_N^2} \nonumber \\
&+& \frac{1}{3}g_M^2(q^2) \frac{q^2}{4m_N^2} ] \nonumber 
\end{eqnarray}
for the $q$ term:
\begin{eqnarray}
f_{qF}&=&j_{1}(qr) qr g_V^2(q^2) \nonumber \\
f_{qGT\pm}&=&f_{qGT,AA}+f_{qGT,AP}+f_{qGT,PP}\mp f_{qGT,MM} \nonumber \\
&=&j_1(qr) qr (g_A^2(q^2)+2g_A(q^2)g_P(q^2)\frac{q^2}{2m_N} \nonumber \\
&-&g_P^2(q^2) \frac{q^4}{4m_N^2} \mp g_M^2(q^2) \frac{q^2}{2 m_N^2}) \nonumber \\
f_{qT\pm}&=&f_{qT,AA}+f_{qT,AP}+f_{qT,PP}\mp f_{qT,MM} \nonumber \\
&=& j_1(qr)qr \frac{1}{3}g_A^2(q^2) - j_1(qr)qr\frac{1}{3} g_A(q^2)g_P(q^2) \frac{q^2}{2m_N} \nonumber \\
&+&\frac{1}{10} (\frac{2}{3}j_1(qr) + j_3(qr)) qr g_P^2(q^2) \frac{q^4}{4m_N^2} \nonumber \\
&\pm& \frac{1}{30} (j_1(qr)+3 j_3(qr)) qr g_M^2(q^2) \frac{q^2}{2m_N^2} \nonumber
\end{eqnarray}
for the $R$ term:
\begin{eqnarray}
f_{RGT} &=& -j_0(qr) q R \frac{q}{3m_N} g_M(q^2) g_A(q^2) \nonumber \\
f_{RT} &=& -j_2(qr) q R \frac{q}{6 m_N} g_M(q^2) g_A(q^2) \nonumber
\end{eqnarray}
and finally for the $P$  term:
\begin{eqnarray}
f_P &=&  j_1(qr) qr_+ g_V(q^2) g_A(q^2) 
\end{eqnarray}
While the radial functions for the $\omega$ term are a bit different:
\begin{eqnarray}
h_I(r)=f_{src}^2(r) \frac{2R}{\pi} \sum_k \int f_{Ik}(q,r)\frac{q^2 dq}{[q+\tilde{E}-(E_i+E_f)/2]^2}
\nonumber \\
\label{omega}
\end{eqnarray}
With 
\begin{eqnarray}
f_{\omega F}&=&j_0(qr) g_V^2(q^2) \nonumber \\
f_{GT\pm}&=& f_{GT,AA}+f_{GT,AP}+f_{GT,PP}\pm f_{GT,MM} \nonumber \\
&=&j_0(qr)[g_A^2(q^2)-\frac{2}{3}g_A(q^2)g_P(q^2)\frac{q^2}{2m_N}\nonumber \\
&+& \frac{1}{3} g_P^2(q^2) \frac{q^4}{4m_N^2} \pm \frac{2}{3}g_M^2(q^2) \frac{q^2}{4m_N^2} ] \nonumber \\
f_{T\pm}&=& f_{T,AP}+f_{T,PP}\pm f_{T,MM} \nonumber \\
&=&j_2(qr)[\frac{2}{3}g_A(q^2)g_P(q^2)\frac{q^2}{2m_N} - \frac{1}{3} g_P^2(q^2) \frac{q^4}{4m_N^2} \nonumber \\
&\pm& \frac{1}{3}g_M^2(q^2) \frac{q^2}{4m_N^2} ] \nonumber 
\end{eqnarray}

The angular part of the matrix element $O_I$ has the forms:
\begin{eqnarray}
\mathcal{O}_{iF} &=& 1 \nonumber \\
\mathcal{O}_{iGT} &=& \vec{\sigma}_{m}\cdot \vec{\sigma}_{n} \nonumber \\
\mathcal{O}_{iT} &=& 3(\vec{\sigma}_m\cdot\hat{r})(\vec{\sigma}_n\cdot\hat{r})-\vec{\sigma}_m\cdot\vec{\sigma}_n \nonumber \\
\mathcal{O}_P &=& i (\vec{\sigma}_{m}-\vec{\sigma}_{n})\cdot (\hat{r}\times\hat{r}_{+})
\end{eqnarray}
Here $i$ refers to different terms such as the $\omega$ term.

With the equations above, we give a complete expression for the NME with weak-current including all LO terms and a sub-leading order (NLO) weak-magnetism term \cite{Cirigliano:2017djv}.

\begin{table*}[htp]
    \centering
    \begin{tabular}{ccc|cc|cc|cc|cc}
    \hline
    \multicolumn{3}{c|}{NME} & \multicolumn{2}{c|}{$^{76}$Ge$\rightarrow ^{76}$Se}& \multicolumn{2}{c|}{$^{82}$Se$\rightarrow^{82}$Kr}&  \multicolumn{2}{c|}{$^{130}$Te$\rightarrow ^{130}$Xe}&  \multicolumn{2}{c}{$^{136}$Xe$\rightarrow^{136}$Ba}\\ \cline{4-11}
         & & &  \quad jun45\quad & \quad jj44b \quad &\quad jun45 \quad &\quad jj44b \quad &\quad jj55a \quad & GCN50:82 &\quad jj55a \quad & GCN50:82 \\
        \hline
 \multirow{10}{*}{$M_{m}$ \quad}& F& &-0.665&-0.601&-0.624&-0.523&-0.668&-0.701&-0.574&-0.567\\
\cline{2-11}
 &\multirow{5}{*}{GT}& AA& 3.584& 3.278& 3.360& 2.860& 3.147& 3.180& 2.648& 2.549\\
 && AP&-1.090&-0.960&-1.021&-0.834&-0.979&-1.034&-0.820&-0.829\\
 && PP& 0.344& 0.300& 0.321& 0.261& 0.313& 0.335& 0.260& 0.268\\
 && MM& 0.247& 0.215& 0.229& 0.188& 0.227& 0.244& 0.188& 0.194\\
\cline{3-11}
 &&total& 3.085& 2.833& 2.889& 2.474& 2.708& 2.724& 2.277& 2.183\\
\cline{2-11}
 &\multirow{4}{*}{T}& AP&-0.013&-0.004&-0.014&-0.012& 0.008& 0.015& 0.002& 0.014\\
 && PP& 0.002&-0.001& 0.003& 0.003&-0.006&-0.007&-0.003&-0.006\\
 && MM&-0.001&-0.000&-0.001&-0.002& 0.003& 0.003& 0.001& 0.002\\
\cline{3-11}
 &&total&-0.012&-0.004&-0.013&-0.010& 0.004& 0.010&-0.000& 0.010\\
\hline
 \multirow{12}{*}{$M_{\omega\pm}$ \quad}& F& &-0.637&-0.575&-0.597&-0.500&-0.637&-0.669&-0.545&-0.540\\
\cline{2-11}
 &\multirow{6}{*}{GT}& AA& 3.276& 2.980& 3.073& 2.596& 2.883& 2.931& 2.427& 2.351\\
 && AP&-1.044&-0.919&-0.978&-0.798&-0.939&-0.993&-0.786&-0.795\\
 && PP& 0.333& 0.290& 0.310& 0.252& 0.303& 0.324& 0.252& 0.259\\
 && MM& 0.239& 0.208& 0.221& 0.181& 0.220& 0.236& 0.182& 0.188\\
\cline{3-11}
 &&GT$_+$total& 2.803& 2.558& 2.626& 2.231& 2.466& 2.498& 2.075& 2.002\\
 &&GT$_-$total& 2.325& 2.172& 2.184& 1.789& 2.026& 2.026& 2.711& 2.626\\
\cline{2-11}
 &\multirow{5}{*}{T}& AP&-0.012&-0.003&-0.013&-0.011& 0.009& 0.015& 0.003& 0.014\\
 && PP& 0.002&-0.001& 0.003& 0.003&-0.006&-0.007&-0.003&-0.006\\
 && MM&-0.001&-0.000&-0.001&-0.002& 0.003& 0.003& 0.001& 0.002\\
\cline{3-11}
 &&T$_+$total&-0.011&-0.004&-0.012&-0.010& 0.005& 0.010& 0.000& 0.010\\
  &&T$_-$total&-0.0013&-0.004&-0.014&-0.014& -0.001& 0.004& -0.001& 0.006\\
\hline
 \multirow{13}{*}{$M_{q\pm}$ \quad}& F& &-0.379&-0.351&-0.359&-0.304&-0.408&-0.417&-0.358&-0.342\\
\cline{2-11}
 &\multirow{6}{*}{GT}& AA& 3.210& 2.981& 3.016& 2.605& 2.781& 2.751& 2.348& 2.209\\
 && AP& 4.842& 4.317& 4.571& 3.741& 4.267& 4.425& 3.607& 3.563\\
 && PP&-1.943&-1.706&-1.829&-1.479&-1.731&-1.827&-1.454&-1.468\\
 && MM&-1.874&-1.636&-1.745&-1.426&-1.708&-1.825&-1.419&-1.456\\
\cline{3-11}
 &&GT$_+$total& 7.983& 7.228& 7.502& 6.293& 7.026& 7.173& 5.920& 5.760\\
  &&GT$_-$total& 4.235& 3.956& 4.012& 3.441& 3.610& 3.523& 3.082& 2.848\\
\cline{2-11}
 &\multirow{6}{*}{T}& AA&-0.056&-0.033&-0.055&-0.042&-0.031&-0.009&-0.031& 0.002\\
 && AP& 0.004&-0.001& 0.006& 0.008&-0.018&-0.018&-0.007&-0.015\\
 && PP& 0.000& 0.001&-0.001&-0.003& 0.007& 0.005& 0.002& 0.003\\
 && MM& 0.000&-0.000&-0.000&-0.001& 0.001& 0.001& 0.000& 0.001\\
\cline{3-11}
 &&T$_+$total&-0.051&-0.034&-0.050&-0.035&-0.043&-0.023&-0.036&-0.012\\
  &&T$_{-}$total&-0.051&-0.034&-0.050&-0.037&-0.041&-0.021&-0.036&-0.009\\
\hline
 \multirow{2}{*}{$M_R$ \quad}& GT &
 & 4.256 & 3.713 & 4.037 & 3.314 & 4.686 & 5.048 & 3.948& 4.080\\
\cline{2-11}
 &T& & 0.014& 0.004& 0.018& 0.028&-0.056&-0.056&-0.014&-0.042\\
\hline
 \multirow{1}{*}{$M_P$ \quad}&  & &-0.431&-0.279&-0.428&-0.152&-0.498&-0.425&-0.289&-0.255\\
\hline
    \end{tabular}
    \caption{$0\nu\beta\beta$-decay NMEs for $^{76}$Ge, $^{82}$Se, $^{130}$Te and $^{136}$Xe from LSSM calculations. Here, for each term we list all the components from different parts as defined in eq.\eqref{NME}. For each nuclei, we present results from two different Hamiltonians as indicated in text.}
    \label{tab:NME}
\end{table*}

\section{Results and Discussion}
\subsection{The NMEs for LR symmetric model}
For the calculations of nuclear matrix elements, we use the large-scale shell model (LSSM) method. For $^{76}$Ge and $^{82}$Se, we use the $jj44$ model space that include four orbitals (0$f_{5/2}$, 1$p_{3/2}$, 1$p_{1/2}$ and 0$g_{9/2}$) for both protons and neutrons. We adopt two Hamiltonians, $jj44b$ \cite{Lisetskiy:2004xp} and $jun45$ \cite{Honma:2009zz}, to understand the possible theoretical uncertainties arising from nuclear interactions within the $jj44$ model space. For $^{130}$Te and $^{136}$Xe, we use the $jj55$ model space that contains five orbitals (0$g_{7/2}$, 1$d_{5/2}$, 1$d_{3/2}$, 2$s_{1/2}$ and 0$h_{11/2}$), and the two Hamiltonians used for this model space are $jj55pn$ \cite{Brown:2004xk} and GCN50:82 \cite{Menendez:2008jp} respectively.

The corresponding NMEs are presented in Table.\ref{tab:NME}. Here, to make comparisons with previous calculations, we divide the NMEs by corresponding functions of $g_A(0)$ or $g_V(0)$. For the neutrino mass mechanism, there are already LSSM calculations for these four nuclei ({\it e.g.} \cite{Senkov:2015juo,Coraggio:2020hwx,Horoi:2013jx,Caurier:2011gi,Menendez:2017fdf}). Our results agree well with previous calculations and we also find that the choice of the closure energy may be the origin of uncertainties, in some cases, leading to a deviation of more than 15\%. In this work, we adopt a closure energy of $7$ MeV. In agreement with results from other calculations, the NME is dominated by the GT part, and the Tensor part is negligible. Meanwhile, the Fermi part is about 1/4 to 1/5 to the GT part, smaller than a naive estimation of $1/3$ using Fierz rearrangement \cite{Prezeau:2003xn} for the short range mechanism case.  As found in the literatures \cite{Brown:2015gsa}, the uncertainties for NMEs with the neutrino-mass mechanism from different Hamiltonians are about 10\%. In our calculation, among the four nuclei, the biggest deviation related the choice of Hamiltonian comes from $^{82}$Se, where the two Hamiltonians lead to an uncertainty of about 20\% for the Fermi part and about 15\% for the GT part. Meanwhile, the ratios of corresponding components inside the GT part from the two Hamiltonians are basically the same. For other nuclei, such uncertainties are limited to less than 10\% for all components.

For other mechanisms, the LSSM calculations are relatively limited, dating back to $90's$ \cite{Caurier:1996bu} and recently some calculations concerning $\lambda$ mechanism \cite{Sarkar:2020cjn,Iwata:2021bgi}. However, systematic studies with nuclear currents including even the weak magnetism term are not yet available. So in this work, we give a thorough analysis of all the relevant long-range terms related to the LR symmetric model.

In the LR symmetric model, different mechanisms exhibit distinct current-current interaction structures. For the nuclear part, both the mass and the $\eta$ mechanisms have left-handed (V-A) nuclear currents at both vertices. However, for the $\lambda$ mechanism, one has the left-handed (V-A) nuclear current at one vertex and right-handed (V+A) nuclear current at another. Given that the weak-magnetism current is induced by the vector current, this implies that for the $\lambda$ mechanism, the MM components has a different sign relative to the other GT components, in contrast to the mass or the $\eta$ mechanism. Therefore, as expressed in eq.\eqref{indNME}, the GT and tensor parts have MM components differed by the sign for the two mechanisms.
 
For the $\omega_+$ term, which has the same nuclear current structure as the mass term, its difference with the mass term comes from the energy denominator [eq.(\ref{omega})].  If we set the closure energy to be zero, then these two terms become identical. With the chosen closure energy, the $\omega$ term is slightly smaller than the mass term. For the F part, we have about 5\% reduction for all cases, irrespective of nuclei or Hamiltonian. For the GT part, the reduction is less than 10\% and is mostly from the AA component, either because it is large in magnitude and also due to the fact that its radial integration is dominated by the low $q$ parts as we will show in subsequent discussions. The MM component is about 10\% in magnitude to the AA component and then the difference between $\omega_-$ and $\omega_+$ terms is about 20\% due to the different signs of the MM component.

The Fermi part is about 1/5 of the GT part for the mass or the $\omega$ terms. Its relative magnitude of the $q$ terms is much smaller, just about 1/10 or less for the two $fp$ nuclei and 1/9 or less for the two heavier ones. Their absolute magnitude is also suppressed, its absolute value is only about 50\% to 60\% of counterparts in the mass and the $\omega$ terms.

A similar situation occurs for the AA component of the GT part. Compared to its counterpart in mass term, it is reduced by more than 10\%. Meanwhile, other components in the $q$ term are significantly enhanced, especially the AP component, which now becomes the dominant contribution to the GT part. Unlike the mass term, the four components accounted for here contribute with similar magnitudes. The major reason is that these components now have similar angular coupling coefficients, in contrast to the mass term. Due to the largeness of the MM component, previous calculations without the inclusion of this weak-magnetism term give an under or over-estimation for $q$ terms \cite{Sarkar:2020cjn,Iwata:2021bgi}. This weak-magnetism component contributes more than half of the AA  component, and less than half of the dominant AP component. This implies that we have to treat it as an equal contribution as other LO terms and include it in future calculations. 
For $q_{+}$ term, the PP and the MM components almost cancels each other. Then, $M_{qGT+}$ is 2-3 times larger than $M_{GT}$ in the mass term. However, for $q_{-}$ term, these two terms nearly cancel contributions from the AP part, allowing for a similar magnitude of $M_{qGT-}$ to $M_{GT}$ in the mass term. 

On the other hand, the tensor part, as in the mass or $\omega$ terms, can be safely neglected. This is common in LSSM calculations while QRPA suggests that the magnitude of tensor part is about 10\%. The reason may stem from the missing orbitals in LSSM compared to QRPA and this still needs further investigation.

For the GT part of the so-called relativistic ($R$-) term \cite{Tomoda:1990rs}, despite the fact that the weak-magnetism current is NLO, its magnitude is of the LO size irrespective of the nucleus or the Hamiltonian. For the two lighter nuclei, $M_{RGT}$ is basically the same as $M_{qGT-}$ and for the two heavier nuclei, it is even larger than the latter. This suggests that although it is counted as NLO because of the factor $q/M_N$, it is actually giving LO contributions mainly due to a large $\mu_p-\mu_n$ \cite{Cirigliano:2017djv}. We will give a more detailed analysis in subsequent sections. For the $R$ term, as other terms, we can safely neglect the tensor part in LSSM calculations. 

The last term to be discussed is the $P$ term. Compared to other terms, it is somehow heavily suppressed due to its special operator structure. In previous calculation~\cite{Caurier:1996bu}, the change of NME signs for different nuclei is observed, but for our calculation, we find for different nuclei, it has always the same negative sign for NME values. While $jun45$ and $jj44b$ gives quite different predictions, the results from $jj55a$ and GCN50:82 are closer, the difference is about 10\%. Meanwhile, our results differs largely from that of LSSM calculation in \cite{Caurier:2011gi} by more than a factor of two, but close to QRPA calculation in \cite{Muto:1989cd}. These discrepancies need further investigations.

\subsection{Enhancement of MM and PP components for the $q$ term}
\begin{figure}
\includegraphics[scale=0.35]{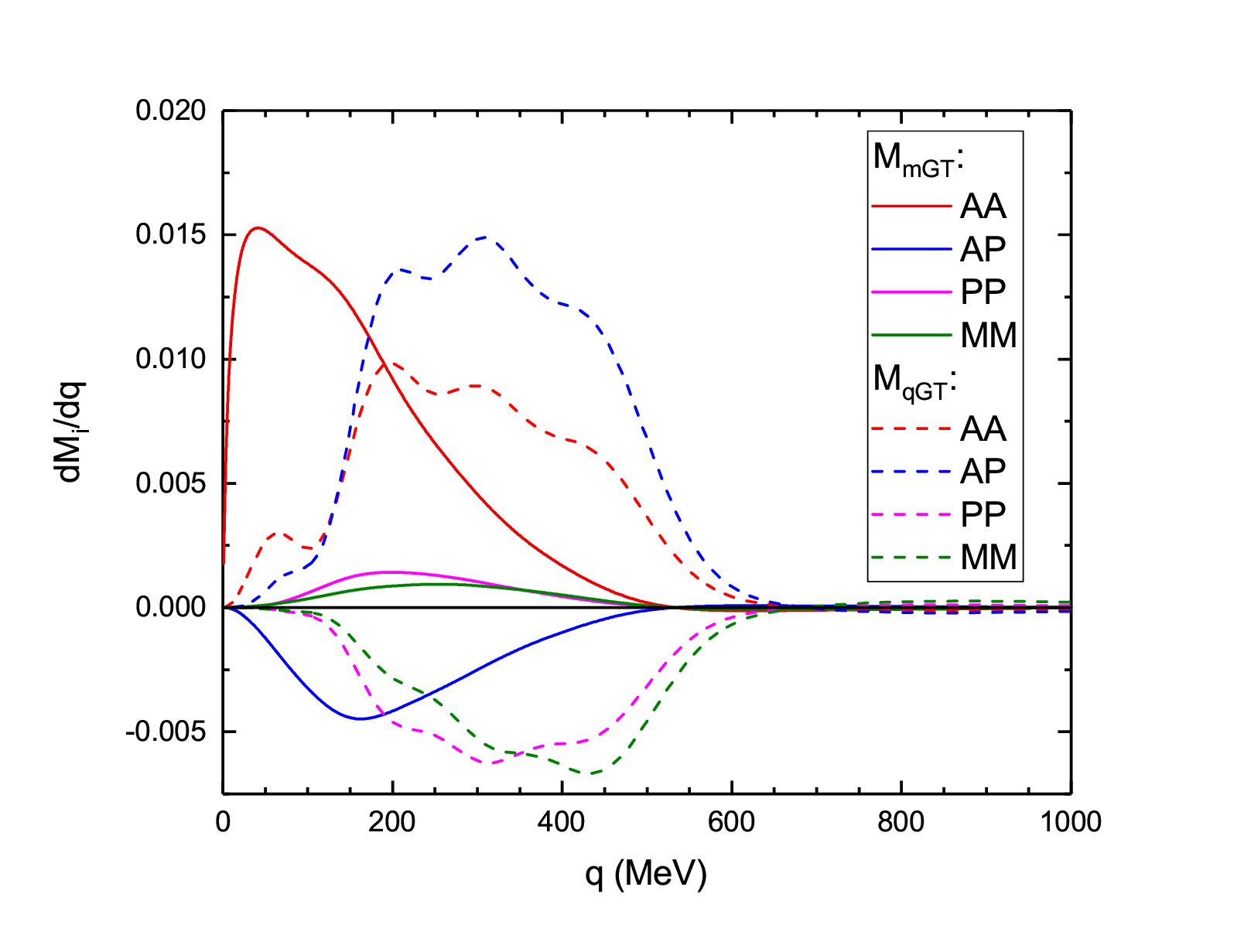}
\caption{(Color online) Strength distribution of different components in $q$ space for $M_{mGT}$ and $M_{qGT+}$ for $^{76}$Ge. For $M_{qGT-}$, the MM part changes the sign.}
\label{qdis}
\end{figure}

The weak magnetic current comes out to be an NLO contribution \cite{Cirigliano:2017djv}. This argument of power-counting is valid if the exchange momentum is smaller or around $m_\pi$. For the NME, the MM component is a scalar product of two weak magnetic currents, it is thus supposed to be suppressed. For the light neutrino mass mechanism, this is the case as MM contributes only several percent in \cite{Simkovic:1999re} and for our calculation. However, for the $q$ term, a different behavior is observed. Although the magnitude of MM component is still smaller than AA or AP as well as PP, its relative ratio to AA is now much larger than that of mass term. Such behavior has already observed for the heavy neutrino mass mechanism \cite{Simkovic:1999re} where the higher exchange momentum dominates in the momentum ($q$) space integration. As seen in above section, the relative ratios of the MM as well as PP components to AA component have been shifted from about 10\% to more than 50\%.

To better understand the large contributions of the MM and PP components, in Fig.\ref{qdis}, we plot the GT NME's strength distribution in $q$ space for both the mass and $q$ terms, for the mass term such studies have already been done \cite{Simkovic:2007vu,Engel:2014pha,Shimizu:2017qcy}. We find that the results in Fig.\ref{qdis} agree with previous studies for mass mechanism, while the behaviors look quite different for the MM parts of these two terms in Fig.\ref{qdis}. The maxima for different components appear at different exchange momenta, this implies that for different mechanisms, the typical exchange momentum is different. For the mass term, it is obvious that the AA component is dominated by the low $q\sim 0\ll m_\pi$ region while AP and PP components are dominated by the region $q\sim m_\pi$. The MM component in this case are also dominated by contributions from the region $q\sim m_\pi$ but is suppressed by a factor of $q/M \sim m_\pi/M$. For the mass term, the overall strength is proportional to the product of the form factor and the spherical Bessel function $j_{0}(qr)$ which has a maximum at $q=0$ , this explains the dominance of the AA component and the relative suppression of other components, especially for the MM component. 

For the $q$ terms, we find that the typical transfer momentum is around $q\sim $400 MeV, this is due to the fact that the integrand is proportional to the spherical Bessel function $j_1(qr)$ which has a node $qr\approx \pi/2$ instead of zero for rank 0 spherical Bessel function $j_0$. In Fig.\ref{qdis}, we can clearly see these nodes of the integrand functions

Therefore, as we have stated above, the naive power-counting may apply to the NME calculations only if one is takes into account of the typical exchange momentum $q$ for the process. For different terms and different components, $q$ may be different, their effective power-counting may differ. 


The discussion can be extended to the analysis of $M_R$, which is the products of axial vector and weak magnetism currents. Compared to the GT matrix elements from mass term, there is an additional factor $q^2$ multiplied to the Spherical Bessel function, this then enhances the contribution from high $q>m_\pi$. When multiplied with an extra factor of $g_M(0)\approx$4.7, we obtain $M_{RGT}$ similar or larger in size as $M_{GT}$, although naively it is expected to be suppressed. 

These analyses suggest that an explicit estimation of the magnitude of NMEs requires our knowledge on the typical exchange momentum for the virtually mediated neutrino. 

\subsection{Estimations of relative magnitudes of different terms}
\begin{table*}[htp]
\caption{A naive estimation of the magnitudes of different terms from $0\nu\beta\beta$-decay rates (see text). For each term, we have two sub rows indicating the results from the two different Hamiltonians: for $^{76}$Ge and $^{82}$Se, the first sub rows are results from jun45 and second sub rows results from jj44b; for $^{130}$Te and $^{136}$Xe, the first sub rows are results from jj55a and the second sub rows from GCN50:82. $r_e$, $r_N$ and $r_R$ are the ratios of square-rooted phase space factors, NMEs and reaction matrix elements of each term to the mass term respectively.}
    \centering
    \begin{tabular}{cc|ccc|ccc|ccc|ccc|ccc}
    \hline
&& \multicolumn{3}{c|}{rough estimation} & \multicolumn{3}{c|}{$^{76}$Ge} & \multicolumn{3}{c|}{$^{82}$Se} & \multicolumn{3}{c|}{$^{130}$Te} & \multicolumn{3}{c}{$^{136}$Xe} \\
\hline
&&lepton & nuclear  &$\mathcal{R}$& $G_{0\nu}$ & $M_{0\nu}$ & & $G_{0\nu}$ & $M_{0\nu}$ & &$G_{0\nu}$ & $M_{0\nu}$ & & $G_{0\nu}$ & $M_{0\nu}$ &\\
\hline
\multirow{2}{*}{$\mu_{\beta\beta}$}& & \multirow{2}{*}{$\mathcal{O}(1)$} & \multirow{2}{*}{$\mathcal{O}(1)$} &\multirow{2}{*}{$\mathcal{O}(1)$}& $0.24$ & 5.62 & & $1.02$ & 5.26 & & $1.43$ & 5.04 & & $1.46$ & 4.25 &  \\
& & & &&& 5.16 && & 4.50 && & 5.11 &&& 4.10 & \\ 
\hline
&&&&&$r_{e}$&$r_N$&$r_{R}$&$r_{e}$&$r_N$&$r_{R}$&$r_{e}$&$r_N$&$r_{R}$&$r_{e}$&$r_N$&$r_{R}$ \\
\hline
\multirow{4}{*}{$\langle\lambda\rangle$}&\multirow{2}{*}{$M_\omega$}&\multirow{2}{*}{$\mathcal{O}(\epsilon_{12}/m_e)$}& \multirow{2}{*}{$\mathcal{O}(1)$} &\multirow{2}{*}{$\mathcal{O}(1)$} & \multirow{2}{*}{1.25} &0.78 & 0.98 & \multirow{2}{*}{1.85} & 0.78 & 1.45 & \multirow{2}{*}{1.61} &0.78 & 1.25 & \multirow{2}{*}{1.57} &0.78 & 1.22 \\
&& & & && 0.78 & 0.98 & & 0.78 & 1.44 & & 0.77 & 1.24 && 0.77 & 1.21 \\
\cline{2-17}
&\multirow{2}{*}{$M_q$}&\multirow{2}{*}{$\mathcal{O}(\omega R)$}& 
\multirow{2}{*}{$\mathcal{O}(q/m_e)$}& \multirow{2}{*}{$\mathcal{O}(1)$} & 
\multirow{2}{*}{0.010} & 55.1 & 0.53 & 
\multirow{2}{*}{0.012} & 54.0 & 0.65 & 
\multirow{2}{*}{0.013} & 44.2 & 0.59 &
\multirow{2}{*}{0.013} & 43.4 & 0.58 \\
&& & & && 53.6 & 0.51 & & 52.6 & 0.63 &  & 43.7 & 0.58 & & 42.5 & 0.57 \\
\hline
\multirow{8}{*}{$\langle\eta\rangle$}&\multirow{2}{*}{$M_\omega$}&
\multirow{2}{*}{$\mathcal{O}(\epsilon_{12}/m_e)$}& 
\multirow{2}{*}{$\mathcal{O}(1)$} & \multirow{2}{*}{$\mathcal{O}(1)$} &
\multirow{2}{*}{1.25} &0.69 & 0.86 & 
\multirow{2}{*}{1.85} & 0.69 & 1.27 & 
\multirow{2}{*}{1.61} &0.66 & 1.07 & 
\multirow{2}{*}{1.57} &0.66 & 1.04 \\
&& & & && 0.69 & 0.86 & & 0.69 & 1.27 & & 0.66 & 1.06 && 0.66 & 1.04 \\
\cline{2-17}
&\multirow{2}{*}{$M_q$}&\multirow{2}{*}{$\mathcal{O}(\omega R)$}& 
\multirow{2}{*}{$\mathcal{O}(q/m_e)$} & \multirow{2}{*}{$\mathcal{O}(1)$} &
\multirow{2}{*}{0.010} & 38.1 & 0.36 & 
\multirow{2}{*}{0.012} & 37.7 & 0.45 & 
\multirow{2}{*}{0.013} & 31.3 & 0.41 &
\multirow{2}{*}{0.013} & 31.4 & 0.42 \\
&& & & && 38.1 & 0.36 & & 37.4 & 0.45 &  & 29.6 & 0.39 & & 29.0 & 0.39 \\
\cline{2-17}
&\multirow{2}{*}{$M_R$}&\multirow{2}{*}{$\mathcal{O}(1)$}& 
\multirow{2}{*}{$\mathcal{O}( q^2/(M_N m_e))$} & \multirow{2}{*}{$\mathcal{O}(\varepsilon^{-1})$} &
\multirow{2}{*}{3.02} & 73.3 & 221.5 & 
\multirow{2}{*}{2.96} & 72.6 & 214.8 & 
\multirow{2}{*}{2.97} & 73.8 & 219.4 &
\multirow{2}{*}{2.97} & 73.5 & 218.4 \\
&& & & && 69.4 & 209.8 & & 70.1 & 207.4 &  & 78.5 & 233.4 & & 77.9 & 231.8 \\
\cline{2-17}
&\multirow{2}{*}{$M_P$}&\multirow{2}{*}{$\mathcal{O}(\alpha Z)$}& 
\multirow{2}{*}{$\mathcal{O}(q/m_e)$} & \multirow{2}{*}{$\mathcal{O}(\varepsilon^{-1})$} &
\multirow{2}{*}{0.34} & 7.40 & 2.49 & 
\multirow{2}{*}{0.33} & 7.65 & 2.50 & 
\multirow{2}{*}{0.27} & 7.97 & 2.19 &
\multirow{2}{*}{0.25} & 5.41 & 1.37 \\
&& & & && 5.21 & 1.75 & & 3.18 & 1.04 &  & 6.71 & 1.84 & & 4.94 & 1.25 \\
\hline
    \end{tabular}
    \label{estm}
\end{table*}

As has been demonstrated in \cite{Doi:1985dx,Tomoda:1990rs}, different terms may be of different relative magnitudes. Thus we adjust the expression of the decay width to better explain the origin of these different magnitudes. A rough estimation for the magnitude for the different terms is given in \cite{Tomoda:1990rs}, suggesting the dominance of $R$ terms and $P$ terms for $\eta$ mechanism, and they also suggest that $\lambda$ mechanism is dominated by $\omega$ term. In this work, we make a more explicit study of the relative magnitudes of various terms for different underlying mechanisms. We start by using the standard mass mechanism as the baseline, since for this term at the leading order, only one term contributes. Therefore, we define the ratios $r_{eI}\equiv \sqrt{\mathcal{G}_I/\mathcal{G}_{01}}$ and $r_{NI}\equiv M_I/M^{0\nu}_{m}$ to denote the relative importance of the lepton and nuclear parts. The detailed values of these ratios are given in Table \ref{estm}, with the rough estimations given at first columns. 

For the nuclear part, we use standard the mass mechanism NME as the baseline. The $\omega$ term $M_{\omega}$ has $\mathcal{O}(1)$. The magnitude of the non-helicity-suppressed NMEs $M^{0\nu}_q$ and $M^{0\nu}_P$ are proportional to $q$, enhanced with a factor $q/m_e$ in our convention. Meanwhile, the relativistic terms \cite{Tomoda:1990rs} $M^{0\nu}_R$ are suppressed by an extra factor of $q/M_N$ from the weak-magnetism term compared to $M^{0\nu}_q$ and $M^{0\nu}_P$, here we neglect the possible contributions from the so-called recoil term \cite{Doi:1985dx,Tomoda:1990rs} which is estimated to be small in~\cite{Tomoda:1990rs}.

For the electron part, the $s$-wave appears at LO [$j_0(kR)\sim\mathcal{O}(1)$], then the $p$-wave function which is connected with $\vec{r}$ operator has a magnitude of $\mathcal{O}(\omega R)$, while the one which is connected with the $\vec{r}_{+}$ operator should have the magnitude $\alpha Z$, much larger than that for operators with $\vec{r}$, this is the so-called $p$-wave effect \cite{Doi:1985dx}. 

To estimate these contributions, we introduce a parameter $\epsilon\approx1/10$ like the $\epsilon_\chi$ from $\chi$EFT introduced in~\cite{Cirigliano:2017djv}, then each of above contributions can be assigned an order. For example, the $\mathcal{O}(q/M_N)$ and $\mathcal{O}(\alpha Z)$ terms appears at the order $\mathcal{O}(\varepsilon)$ while $\mathcal{O}(\omega R)$ and $\mathcal{O}(m_e/q)$ appears at the order of $\mathcal{O}(\varepsilon^2)$. 


A detailed analysis based on our numerical results for NME and previous numerical results for PSFs in \cite{Stefanik:2015du} are presented in Table \ref{estm}. Starting from the $\omega$ term, the electron part is basically the same as the mass term, and so does the NME, agree with the assigned order $\mathcal{O}(1)$ in Table. \ref{estm}. Its contribution to the $\lambda$ mechanism is generally larger than that to the $\eta$ mechanism due to opposite contributions from the Fermi part.  For the q terms, with assigned order $\mathcal{O}(1)$, ratios of about 1/2 to standard mechanism for $\lambda$ mechanism and 1/2 to 1/3 for $\eta$ mechanism are observed, slightly smaller than the naive estimation due to a suppression from the nuclear part. The $p$-wave effect makes the electron part of the $P$ term more than one order of magnitude larger than that of the $q$ term. Meanwhile, $P$ term's NME is heavily suppressed, order of magnitude smaller than expected. This is actually observed by LSSM calculations earlier \cite{Caurier:1996bu}. Therefore, even with the so-called $p$-wave effect~\cite{Doi:1985dx}, the $P$ term gives an $\mathcal{O}(1)$ contribution instead of the expected $\mathcal{O}(\varepsilon^{-1})$ in Table \ref{estm}. On the other hand, because of the enhanced NMEs, the $R$ term gives an $\mathcal{O}(\varepsilon^{-2})$ contribution instead of the $\mathcal{O}(\varepsilon^{-1})$, one order of magnitude larger than naive estimation. This makes the $R$ term dominate the $\eta$ mechanism, while other terms two orders of magnitude smaller. 

Therefore, without considerations of the new physics parameters, we need to slightly modify the naive estimation of orders of contributions. From above results, we find that $R$ has a much larger size than others, it is natural to assume $R$ term to be at the leading order $\mathcal{O}(1)$, the $P$-term which is supposed to be the same size as $R$ term is now suppressed by the nuclear part and comes out to be of the similar size as $\omega$ term, they all appear at the order $\mathcal{O}(\varepsilon^2)$, so does the mass term. Meanwhile, although the $q$ term is suppressed, it can still be account as $\mathcal{O}(\varepsilon^2)$. 

In general, for LSSM calculations, we find that the $\eta$ mechanism is dominated by the $R$ term, while corrections from other terms are at the percent level (at the order of $\mathcal{O}(\varepsilon^2)$). For the $\lambda$ mechanism, the $\omega$ term at the order of $\mathcal{O}(\varepsilon^2)$ dominates but receives an additive correction of about 50\% from the $q$ term. Now, if we incorporate the new physics parameters, we find that if these different mechanisms coexist, then one requires that $\langle \lambda \rangle \sim \mu_{\beta\beta}\sim (m_e/q) \langle \eta \rangle$. This analysis helps to constrain the new physics model, and we will proceed to this topic in the next section.   

\subsection{Current experimental constraints on LR-symmetric model}

Current lower limits on half-lives of neutrinoless double beta decay collected from various literatures are: $1.8\times10^{26}$yr at 90\% C.L. for $^{76}$Ge from GERDA \cite{GERDA:2020xhi}, $4.6\times10^{24}$yr at 90\% C.L. for $^{82}$Se from CUPID \cite{CUPID:2022puj}, $2.2\times 10^{25}$yr at 90\% C.L. for $^{130}$Te from CUORE \cite{CUORE:2021mvw}, and $2.3\times 10^{26}$yr at 90\% C.L. for $^{136}$Xe from KamLAND-Zen \cite{KamLAND-Zen:2022tow}. The most stringent constraint is from KamLAND-Zen, while GERDA also provide very tight constraints larger than $10^{26}$yr. Assuming that the mass term dominates and set $g_A=1.27$, we obtain the most stringent constraint on the effective neutrino mass for $m_{\beta\beta}<66$ meV from $^{136}$Xe with the $jj55a$ Hamiltonian in our calculation, these results are approaching the so-called inverted hierarchy mass region. However, there are results with extremely small NMEs from other many-body approaches \cite{Mustonen:2013zu,Fang:2018tui} for this nucleus, which may push the effective neutrino mass limit to a higher value region.

\begin{figure*}
\includegraphics[scale=1.0]{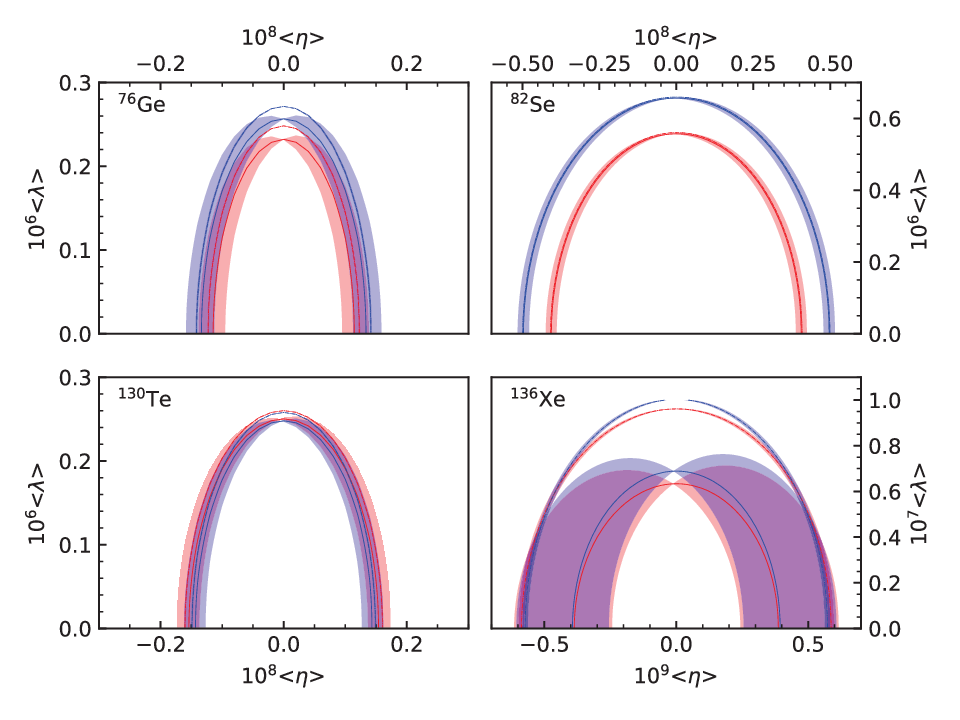}
\caption{(Color online) Constraints on the LR symmetric model from current experimental limit\cite{GERDA:2020xhi,CUPID:2022puj,CUPID:2022puj,KamLAND-Zen:2022tow}.Here the we take the double beta decay effective neutrino mass $m_{\beta\beta}$ with two typical values 50meV(solid lines) and 3 meV(dashed lines) corresponding to IH and NH respectively. Here the shaded area comes from the uncertainties of the phase angle $\psi$'s with the assumption $\psi_1=\psi_2$. The different colors correspond to different Hamiltonians one adopts: red for $jj44b$ and blue for $jun45$ for $^{76}$Ge and $^{82}$Se; red for $jj55a$ and blue for GCN50:82 for $^{130}$Te and $^{136}$Xe. }
\label{constr}
\end{figure*}

Current experiments still leave space for the coexistence of different mechanisms if the neutrino mass indeed has an inverted hierarchy. Therefore, in Fig.\ref{constr}, we choose typical effective neutrino masses and then provide the constraints for $\lambda$ and $\eta$. With current LR symmetric model, $\langle \lambda \rangle$ is defined as positive, and $\Psi_1=\Psi_2$ leaves only one uncertain phase angle. We chose two typical effective neutrino masses of $50$ meV and $3$ meV, which lie in the inverted and normal hierarchy neutrino mass regions, respectively.

We first consider the inverted hierarchy case. The relatively large $m_{\beta\beta}$ leaves a smaller room for the parameter space of $\langle \lambda \rangle$ and $\langle \eta \rangle$. The most stringent constraints on $\langle \lambda \rangle$ is from $^{136}$Xe, which requires that $\langle \lambda \rangle$ be smaller than about $6-7\times10^{-8}$ depending on both the Hamiltonian and the phase angle. The uncertainties (shaded area) caused by the phase angle are related to the interference terms {(the second and third terms in r.h.s. of eq.\eqref{LRDW} )} and related to the magnitudes of $\mathcal{C}_{m\lambda}$ and $\mathcal{C}_{m\eta}$, the latter determines the width of the uncertainty band. Meanwhile the most stringent constraints on $\langle \eta \rangle$ from our calculations come also from kamLAND-Zen, it suggests that $|\langle \eta \rangle|< 2-7\times 10^{-10}$, smaller than the constraints on $\langle \lambda \rangle$. This can be explained by rough analysis of magnitudes in previous sections, that the contributions for $\langle \eta \rangle$ come out at LO dominated by $R$ term while that for $\langle \lambda \rangle$ come out at next to sub leading order (N2LO). For the other three nuclei, current constraints on $\langle \eta \rangle$ could be one order of magnitude larger due to shorter half-life limits. Especially for $^{82}$Se, half-life limit needs to be improved in the near future from various experiments proposed~\cite{NvDEx:2023zht} to strengthen the constraints.

For a typical normal hierarchy case, we set $m_{\beta\beta}=3 $meV then vary $\langle \eta \rangle$ and $\langle \lambda \rangle$. In Fig.\ref{constr}, the most stringent constraint is again from $^{136}$Xe, although now the constraints are a bit looser than that of the inverted hierarchy case, the difference turns out to be within a factor of two. Now the upper bounds of $\langle \lambda \rangle$ is around $10^{-7}$, and the upper bound for $|\langle\eta\rangle|$ is a bit larger than $6\times10^{-10}$. The uncertainty caused by the phase angle is now much smaller than at the inverted hierarchy case due to the smallness of $\mu_{\beta\beta}$ providing $C_{m\eta}$ and $C_{m\lambda}$ don't change. Also, more stringent constraints on $\langle \eta \rangle$ are due to the fact that the corresponding amplitude is at a lower order as mentioned above because the NME for the $R$ term is enhanced making it a LO contribution. Improved oscillation experiments on the neutrino mass and their hierarchy will provide tighter constraints on these new physics parameters. 
With Fig.\ref{constr}, {we can see the uncertainty from NME arising from our choices of Hamiltonians} on the determination of new physics parameters. 

\section{Conclusion and Perspective}
In this work, we calculate the NMEs of the neutrinoless double beta decay process in the LR symmetric model for four nuclei ($^{76}$Ge, $^{82}$Se, $^{130}$Te, and $^{136}$Xe) using the LSSM approach. We find that the weak magnetism component plays a much more important role in the $R$ term of the $\eta$ mechanism and the $q$ term of the $\lambda$ mechanism than expected. We compare the relative magnitude of each term and discover that if the three new physics parameters have a similar size, the relativistic term becomes the dominant one while the $p$-wave effect is suppressed due to the smallness of the $P$ term's NME. Based on our nuclear structure calculations, combined with previous phase space factor calculations, we provide the constraints on the three new physics parameters under current experimental limits. Future calculations using more many-body approaches are needed for calibrating the NME.

\section*{Acknowledgement}
This work is supported by National Key Research and Development Program of China (2021YFA1601300).This work is also supported by Chinese Academy of Sciences Project for Young Scientists in Basic Research (YSBR-099) and the "from 0 to 1" innovation program ZDBS-LY-SLH014. B.A.B is supported by NSF Grant PHY-2110365. F.\v{S}. acknowledges support from the Slovak Research and Development Agency under Contract No. APVV-22-0413, VEGA Grant Agency of the Slovak Republic under Contract No. 1/0418/22 and by the Czech Science Foundation (GAC\v{C}R), project No. 24-10180S. We thank Prof. M. Horoi for the helps on the TBTD calculation.
\bibliography{LR}
\end{document}